\documentstyle[graphicx]{jaa}

\begin{document}

\title[Radio variability of the first 3-months $\it{Fermi}$ blazars at 5 GHz]{Radio variability of 1st 3-months $\it{Fermi}$ blazars at 5 GHz:
 affected by interstellar scintillation?}
\author[X. Liu et al.]%
       {X. Liu$^1$\thanks{e-mail:liux@uao.ac.cn}, Z. Ding$^{1,2}$, J. Liu$^1$, N. Marchili$^3$, T.P. Krichbaum$^3$\\
        $^1$ Urumqi Observatory, NAOC, 40-5 South Beijing Road, Urumqi 830011, P.R. China\\
        $^2$ Graduate University of Chinese Academy of Sciences, Beijing 100049, P.R.
        China\\
        $^3$ Max-Plank-Institut f\"ur Radioastronomie, Auf dem H\"ugel 69, 53121 Bonn, Germany}


\maketitle

\label{firstpage}

\begin{abstract}
Blazars from the first-three-months $\it{Fermi}$-AGN list were
observed with the Urumqi 25\,m radio telescope at 5\,GHz in IDV
(Intra-Day Variability) mode and inter-month observation mode. A
significant correlation between the flux density at 5\,GHz and the
$\gamma$-ray intensity for the $\it{Fermi}$-LAT detected blazars is seen.
There is a higher IDV detection rate in $\it{Fermi}$ detected
blazars than those reported for other samples.
Stronger variability appears at lower Galactic latitudes; IDV appears to be
stronger in weaker sources, indicating that the variability is affected by
interstellar scintillation.

\end{abstract}


\section{Introduction}

In the 1990s the space $\gamma$-ray telescope EGRET identified 66
blazars. $\it{Fermi}$, launched in 2008, has a much higher
sensitivity and pointing accuracy than EGRET. After first
three months of observations with the $\it{Fermi}$-LAT (Large Area
Telescope) 132 bright $\gamma$-ray sources are detected, of which 104
are blazars (Abdo et al. 2009).

Blazars are either flat-spectrum radio quasars or BL Lacerate
objects, and they are extremely variable at all wavelengths on
timescales ranging from less than an hour to many years. Such
violent behavior in blazars is attributed to relativistic jets
oriented very close to our line of sight (Urry \& Padovani 1995).

Intra-Day Variability (IDV, fast variability on time scales from
a few hours to few days) of the radio flux density has been found in about
30--50\% of all flat-spectrum radio sources (Quirrenbach et
al. 1992; Lovell et al. 2007). If interpreted as being source
intrinsic, the rapid variability would imply micro-arcsecond scale
sizes of the emitting regions. Alternatively, IDV can be caused by
interstellar scintillation (ISS), especially for very rapid
variability seen in some sources, e.g. Gabanyi et al. (2007).

Since most of the $\it{Fermi}$-LAT detected AGN are blazars, it is expected that
$\gamma$-ray emitting AGN are more variable than non $\gamma$-ray emitting AGN.
Therefore we launched in March 2009 a 5\,GHz monitoring program
with the Urumqi 25\,m radio telescope, to observe the intra-day variability and
the inter-month variability of first three-months detected $\it{Fermi}$
blazars with declination $>$0$^\circ$. We aim to enlarge the number of known
IDV sources and to study the statistical occurrence of (rapid)
variability in $\it{Fermi}$ blazars.

\section{Observations and data reduction}

The IDV observations were carried out from March to May 2009 with a
duration of 4-6 days per session. The sources were observed typically every 3
hours using the `cross-scan' method. Sources which were not point-like for
the Urumqi beam or were confused (non Gaussian brightness profiles) were
rejected from our sample. Finally, we link our observations to the
absolute flux density scale of Ott et al. (1994). From the variability
light curves from each source, we obtain the modulation index
($m$), and the mean flux density ($<S>$).

Following Kraus et al. (2003), $m$ is defined by the ratio of the standard
deviation of the flux density and the mean flux density of source.
It provides a reliable measure of the strength of the observed
amplitude variations:

\begin{displaymath}
m[\%]=100\frac{\sigma_{S}}{<S>}
\end{displaymath}
The modulation index $m_{0}$  observed for the calibrator sources provide
a measure for the residual calibration errors, which are around 0.5 [\%].

The inter-month observations were carried out from March to
December 2009, with one flux density measurement per month for
most sources. We define the modulation index $M$ for the
inter-month flux density variations similar as $m$ for the IDV.
For the calibrators we obtain $M_0 \sim$2 [\%].  Finally, 42
blazars are observed in the IDV and inter-month observations; 11
blazars are not included in the IDV and inter-month observations,
most of them are relatively weak with flux densities measured in
March 2009.

\section{Results and discussion}

A significant correlation between the flux density at 5 GHz and
$\gamma$-ray intensity in the $\it{Fermi}$-LAT blazars has been
found and is shown in Fig.1 left, with Pearson correlation coefficient
of 0.42 in total, and 0.40 for QSOs and 0.59 for BL Lacs. Following a $\chi^{2}$-test
for variability, we find 26 sources (16 QSOs and 10
BL Lacs) showing intra-day variability at a confidence level of
larger than 99.9\%, from 42 sources. The IDV detection rate of 62\,\% in the $\it{Fermi}$ blazar
sample is higher than those reported in previous flat-spectrum AGN
samples. This could be caused by a higher compactness
of $\it{Fermi}$ blazars relative to sources in other samples.

The majority of the sources also show inter-month variability, no
obvious correlation was found between the intra-day and
inter-month variability. For the still
very short time coverage of the inter-month data, one may need a much
longer time coverage for a better characterisation of the inter-month variability. 
The median values of the variability
strength for $m$ or $M$ are quite similar for quasars and BL Lacs.
Pronounced inter-month variability ($\sim$40\%) was found in two
BL Lac objects: B\,0109+224 and B\,0235+164. There is only a weak or
marginal correlation between the source spectral index and either the
intra-day or inter-month variability index, since all sources exhibit
similar flat spectra.

On the other hand, in our sample stronger intra-day variability
($m$) appears in weaker sources as shown in Fig.\,1, right. This
could be explained by ISS as suggested by Lovell et al.
(2007). Another interpretation of this phenomenon is that the weak ones might show less powerful and
less pronounced jets. From VLBI we know that the IDV comes from the core region, 
so a 'naked' core might show a higher variability index than a source which shows a long 
and prominent jet. It is well possible that faint sources show less jet emission, which means 
that they are more core-dominated. This would automatically lead to more pronounced IDV. 
Our data show stronger inter-month variability (larger
$M$) at lower Galactic latitude except for two outliers B\,0109+224 and
B\,0235+164, as shown in Fig.2 left. This also points towards
interstellar scintillation similar to the effect previously found by
Rickett et al. (2006) in data from the Green Bank Interferometer. However,
we should check if the sky coverage of our sample is sufficiently uniform.
Fig.\,2 right reveals some deficiency of our sample at $50^{\circ}-80^{\circ}$
latitude. A future extension of the size of the source sample and a
longer time coverage for the long-term variability should help to discriminate
between propagation induced effects (ISS) and source intrinsic properties.
Following the pilot study presented in this paper, we now have launched
a program `Search for rapid variability in a large sample of radio sources
with the Urumqi telescope' in 2010. As parent sample, we use the
CRATES (Healey et al. 2007) catalogue to study the variability statistics in more detail.
\\

\noindent
This work is supported by the National Natural Science Foundation
of China (No.10773019 and 11073036) and the 973 Program of
China (2009CB824800). We acknowledge A. Readhead for discussion
on the manu\-script.


\begin{figure}
     \centering
     \includegraphics[width=0.48\textwidth]{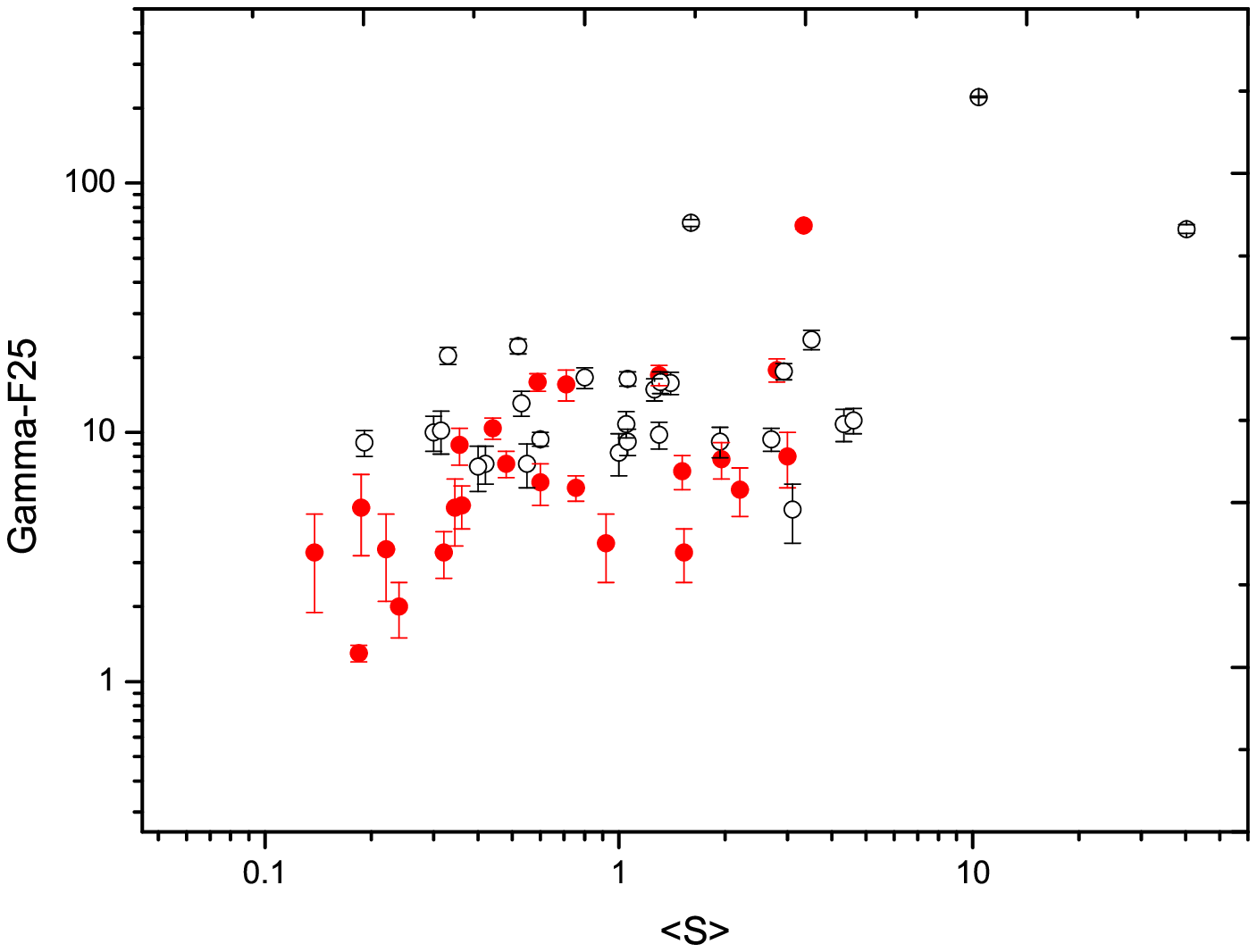}
     \includegraphics[width=0.48\textwidth]{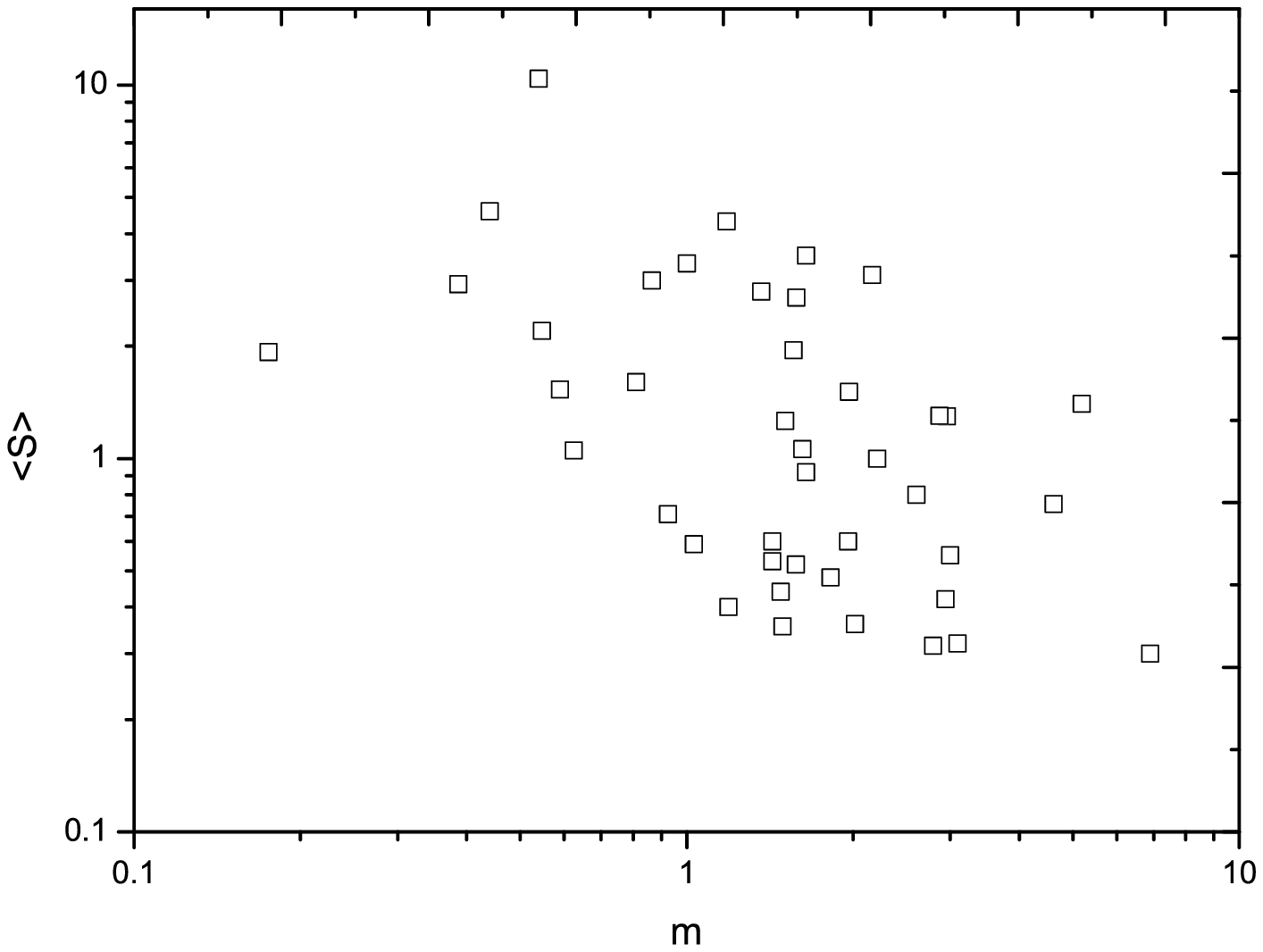}
     \caption[]{Left: The $\gamma$-ray intensities (with error)
     ($\geq100$ MeV, in units of $10^{-8}ph\, cm^{-2}\, s^{-1}$) for 53 $\it{Fermi}$ detected blazars plotted versus the flux densities (Jy) at 5 GHz, filled: for BL Lacs, circles: for QSOs. Right: The mean flux densities (Jy) plotted versus the modulation indices of 42 $\it{Fermi}$ detected blazars from our IDV observations.}
\label{Fig1}
   \end{figure}

\begin{figure}
     \centering
     \includegraphics[width=0.48\textwidth]{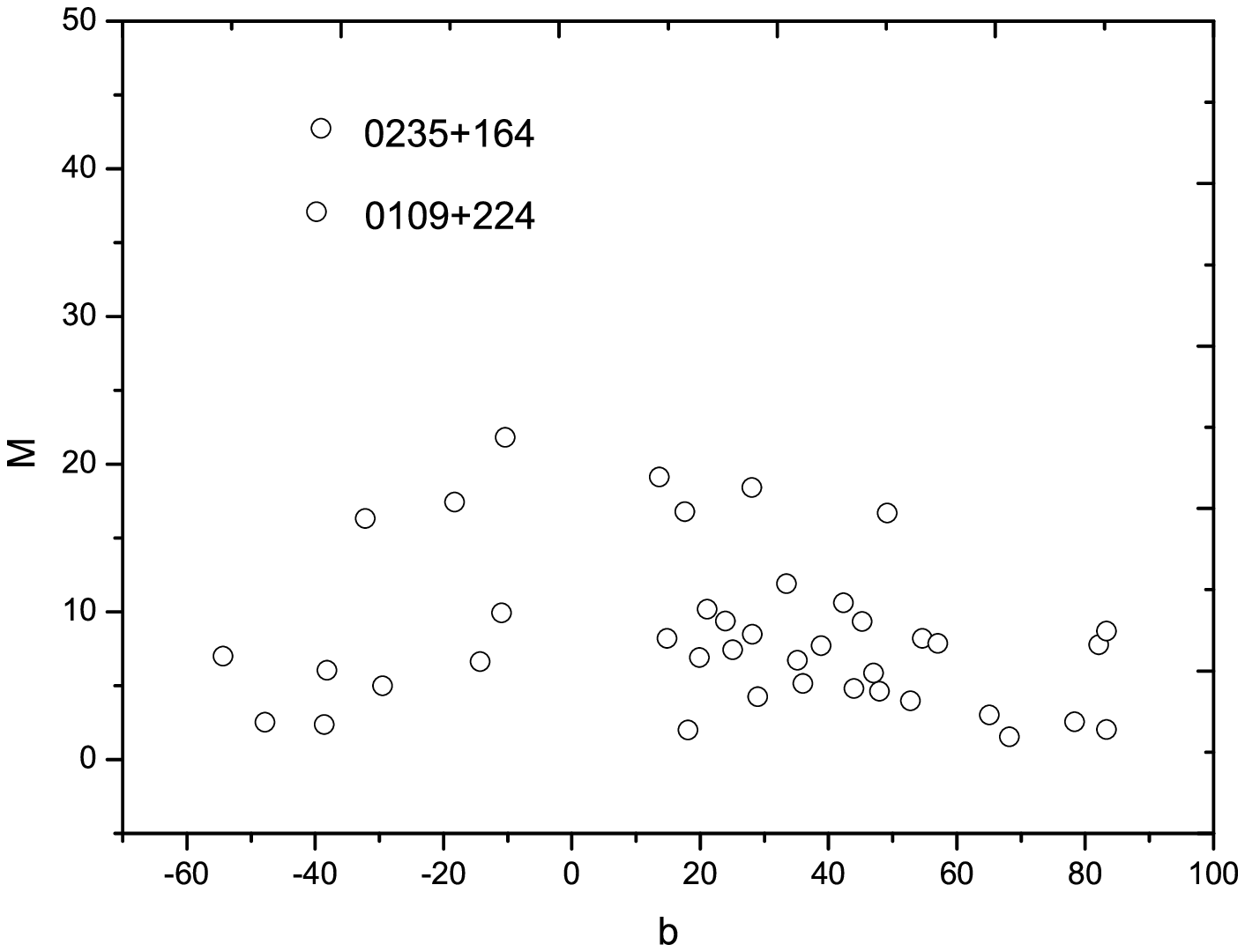}
     \includegraphics[width=0.48\textwidth]{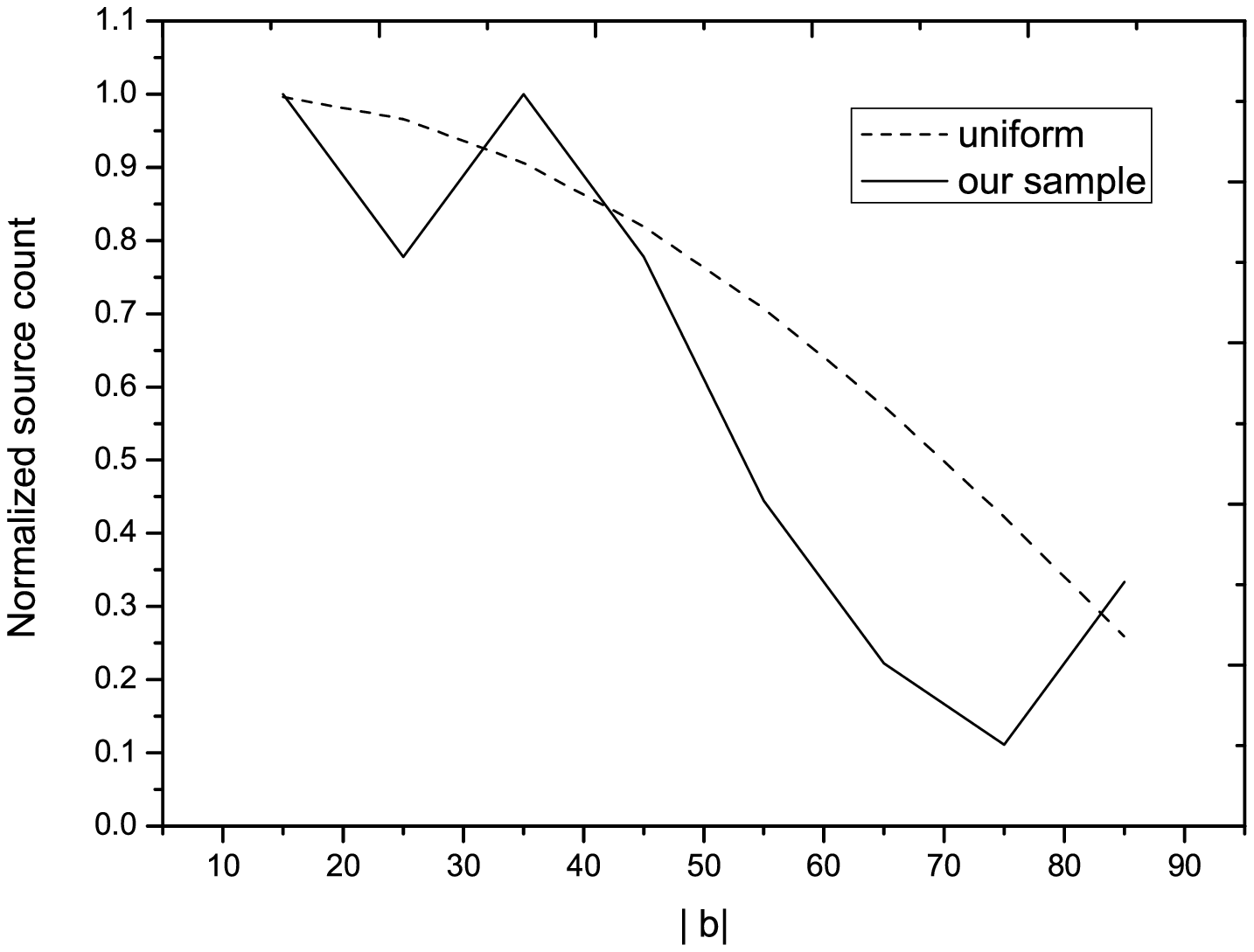}
     \caption[]{Left: The modulation index of inter-month variability versus Galactic latitude of the source. Right: Normalized source count of our sample (solid line) and that for a uniform sky distribution (dashed  line) versus Galactic latitude $| b|$.}
\label{Fig2}
   \end{figure}

\end{document}